\documentclass[12pt]{article}

\usepackage[margin=1in]{geometry}
\usepackage{setspace}
\usepackage[T1]{fontenc}
\usepackage[utf8]{inputenc}
\usepackage{lmodern}        
\usepackage{microtype}
\usepackage{sgame}
\usepackage{tikz}
\usepackage{pgfplots}
\usetikzlibrary{arrows.meta,positioning,calc}
\pgfplotsset{compat=1.18}
\usetikzlibrary{intersections}
\usepgfplotslibrary{fillbetween}

\pgfmathdeclarefunction{cutoffroot}{3}{%
  \pgfmathparse{
    (
      (1-#1)*(#2*#1-(1-#1))
      + #3*sqrt(
        ((1-#1)*(#2*#1-(1-#1)))^2
        + 4*#1*(1-#1)*(#1*(1-#2*(1-#1)))
      )
    )/(2*#1*(1-#1))
  }%
}
\pgfmathdeclarefunction{clearcoop}{3}{%
  \pgfmathparse{#1+(1-#1)*cutoffroot(#1,#2,#3)}%
}
\pgfmathdeclarefunction{youngrate}{3}{%
  \pgfmathparse{
    (1-#1+#1*clearcoop(#1,#2,#3))*clearcoop(#1,#2,#3)
  }%
}

\usepackage{amsmath,amssymb,amsthm,mathtools,bm}

\usepackage{graphicx}
\usepackage{booktabs}
\usepackage{caption}
\usepackage{subcaption}
\usepackage{float}

\usepackage{enumitem}
\usepackage{xcolor}

\usepackage[round,authoryear]{natbib} 
\usepackage{hyperref}
\hypersetup{
  colorlinks=true,
  linkcolor=blue,
  citecolor=blue,
  urlcolor=blue
}
\usepackage{cleveref} 

\numberwithin{equation}{section}
\newtheorem{theorem}{Theorem}[section]
\newtheorem{proposition}[theorem]{Proposition}
\newtheorem{lemma}[theorem]{Lemma}

\theoremstyle{definition}

\theoremstyle{remark}

\title{Honesty, Stigma, and Cooperation in an Overlapping-Generations Game\footnote{We thank Ronaldo Carpio, the editor, and an anonymous referee for helpful comments and suggestions. Georgy Lukyanov acknowledges support from the French National Research Agency (ANR) under grant ANR-17-EURE-0010 (Investissements d'Avenir programme).}}

\author{David Li\footnote{School of Economics, Peking University, 5 Yiheyuan Road, Haidian District, Beijing, P.R.China. Email address: dshli@edu.hse.ru} \and Georgy Lukyanov\footnote{Toulouse School of Economics, 1, Esplanade de l'Université 31080 Toulouse, France. E-mail address: georgy.lukyanov@tse-fr.eu}}

\date{}

\begin{document}

\maketitle

\begin{spacing}{1}
\begin{abstract}
We study a two-period overlapping-generations game in which a young agent interacts with an old agent and carries a binary record into old age. Honest agents cooperate with clear partners, whereas strategic agents have a privately known loss from being exploited. A clear record is informative: honest agents remain clear, while some strategic agents acquire stigma. We incorporate this selection into Bayes-consistent stationary equilibrium beliefs. With uniformly distributed losses, strategic young agents use a cutoff rule and equilibrium reduces to a quadratic equation. A simple parameter region has a unique equilibrium; when the temptation to defect is high, low- and high-cooperation corner equilibria coexist with an interior tipping equilibrium. Along that interior branch, a larger honest share lowers both the strategic cutoff and the young cooperation rate. The branch is locally unstable, however, so the result describes a moving coordination threshold rather than a deterioration of stable equilibrium selection. Under a maximin decision rule, strategic cooperation disappears. Probabilistic record clearing weakens incentives at every fixed conjecture, reduces cooperation at locally stable interior equilibria, and under full clearing leaves only defection by strategic agents.
\end{abstract}


\end{spacing}

\section{Introduction}\label{sec:intro}

Cooperation often has to survive a change of partners: a junior supplier deals with an established buyer, a new professional works with a senior member of another organization, or a first-time platform participant is matched with an incumbent.\footnote{Throughout, the record is \emph{directed} in a specific sense: it is created by one match and consulted by exactly one later match, never broadcast to the community at large. This is more restrictive than the community-wide or stationary record structures usual in the literature on bounded and public memory; Section~\ref{sec:literature} situates the comparison.} The relationship may be short, but conduct within it can affect the terms on which the junior is treated once she becomes the incumbent. This creates an intergenerational reputational incentive even though no pair ever meets twice. How much discipline can a single directed bit of memory extract from agents who would otherwise have every reason to exploit a partner they will never see again---and is a larger share of honest commitment types always good news for cooperation?

We study this question in a deliberately sparse environment. Each agent acts once when young and once when old. A young agent who defects against a cooperating old opponent receives a binary stigma that is observed by the agent's next partner. Some agents are honest commitment types: they cooperate with a clear opponent and defect against a stigmatized one. The remaining agents are strategic and differ privately in the loss they suffer when they cooperate against a defector. Strategic old agents defect because they have no future interaction. Strategic young agents may nevertheless cooperate in order to enter old age with a clear record.

The label changes not only behavior but also beliefs. Starting from clear records, honest agents never become stigmatized, whereas strategic agents can. Consequently, an old agent with a clear record is more likely to be honest than a randomly selected member of the population. Our equilibrium analysis accounts for this selection explicitly: a Perfect Bayesian treatment of the model requires it, and doing so changes the fixed-point equation relative to one that simply keeps the honest probability at its unconditional value.

We show that stationary symmetric cutoff equilibria admit a full characterization once losses are uniformly distributed. Bayes' rule turns the indifference condition into a quadratic equation, with exact conditions for the zero- and unit-cutoff corners and a closed-form expression for every interior cutoff. If the temptation payoff is no larger than the golden ratio, the equilibrium is unique. If it is larger than two, there is a transparent parameter region with three equilibria: two strict corners and an interior cutoff between them.

The interior branch is where intuition needs the most care. Common sense offers two guesses about what a larger honest share should do to cooperation, and neither survives intact. The first guess is that more honesty can only help: if opponents are more often trustworthy, cooperation should rise. The second guess, once the interior comparative static is in view, overcorrects in the other direction and treats honesty as actively harmful. What is shown instead is narrower, and more interesting than either. Within the three-equilibrium region, a larger share of honest agents does lower the interior cutoff, the probability of cooperation with a clear opponent, and the unconditional cooperation rate \emph{along that branch}. But the branch is the tipping point separating the two strict corner equilibria, not a stable outcome in its own right, and a lower tipping point enlarges the set of conjectures from which the high-cooperation corner is reached. The comparative static is a fact about where the boundary sits, not an unconditional verdict on whether honesty helps or hurts.

We then examine two robustness questions. First, we compare Bayesian expected-payoff choice with a maximin rule. A strategic young agent who evaluates each action at the worst possible current opponent type always defects: the worst-case payoff from cooperation is lower by the private exploitation loss, for every strictly positive loss. Second, we allow a newly created stigma to be cleared before the next match. Clearing makes defection more attractive at every fixed conjecture: it lowers cooperation at locally stable interior equilibria and raises an unstable tipping cutoff, and if records are always cleared, every strategic young agent defects. A short extension shows that discounting weakens reputational incentives through the same channel, provided the current temptation payoff is not mistakenly discounted along with the future one.

The paper is related to, but distinct from, our earlier work \citep{LukyanovLi2025}. Both papers use honest and strategic types and private losses to obtain cutoff behavior. The earlier paper is a static two-player Prisoner's Dilemma and asks how common versus heterogeneous beliefs about a partner's honesty affect cooperation. The present paper instead has common knowledge of the primitive honest share, two-period lives, an endogenous record distribution, Bayes updating from records, and a reputational link between successive partners. Neither overlapping generations nor record-based community enforcement appears in the earlier paper: the question there is what agents believe about a fixed partner, while the question here is what a population can sustain once beliefs are pinned down by an endogenous, one-bit history. These differences define the contribution of the present paper.

The remainder of the paper proceeds as follows. Section~\ref{sec:literature} relates the model to the literature. Section~\ref{sec:model} presents the environment, terminology, and applications. Section~\ref{sec:equilibrium} characterizes stationary equilibrium. Section~\ref{sec:comparative} studies cooperation and stability. Section~\ref{sec:robust} gives the maximin benchmark. Section~\ref{sec:forgiveness} studies record clearing and discounting. Section~\ref{sec:conclusion} concludes.

\section{Related literature}\label{sec:literature}

Community-enforcement models show how information about past conduct can support cooperation when partners are randomly matched. Labels and social norms can transmit punishments beyond a bilateral relationship \citep{Kandori1992,Ellison1994}. More recent work establishes both folk-theorem and anti-folk-theorem results in anonymous matching environments \citep{DebSugayaWolitzky2020,SugayaWolitzky2020}. Our agents are not long-lived participants in an indefinitely repeated game. They have only one young and one old interaction, and a one-bit record links those interactions.

The record structure connects the paper to work on bounded memory and the design of reputational histories. Coarse or finite records can be essential for community enforcement \citep{BhaskarThomas2019}, while the updating rule and the amount of information retained determine which cooperative outcomes can be sustained \citep{ClarkFudenbergWolitzky2021}. Record deletion can also shape experimentation and trade \citep{KovbasyukSpagnolo2024}, and agents' ability to manipulate records introduces additional incentive constraints \citep{Pei2026}. Our contribution is narrower: we solve a two-period model in which a directed record is seen only by the next partner, and we make explicit the selection of types induced by that record.

Overlapping-generations games study how incentives and information survive cohort turnover \citep{LagunoffMatsui2004}. The alternating two-group structure also recalls models of intergroup interaction such as \citet{AcemogluWolitzky2014}, although our stage game, individual labels, and source of dynamics are different. Related reputation models show how an individual or firm can carry incentives across career stages \citep{Tadelis2002,BoardMeyerTerVehn2013}. Here the record is not traded and has value for only one subsequent interaction.

Finally, honest agents are commitment types in the tradition of \citet{KrepsMilgromRobertsWilson1982} and resemble the intrinsically rule-following agents discussed by \citet{Frank1987}. Evolutionary models provide one foundation for moral preferences \citep{AlgerWeibull2013}; we instead take the honest share as a primitive. Section~\ref{sec:robust} also relates the Bayesian solution to maximin choice under ambiguity \citep{GilboaSchmeidler1989}.

\section{Model, terminology, and applications}\label{sec:model}

\subsection{Demography, matching, and records}

Time is discrete. There are two populations, $A$ and $B$, each containing a continuum of agents of unit mass in every cohort. An agent belongs to the same population throughout life, is young at date $t$, and is old at date $t+1$. At each date, every young agent from $A$ is matched uniformly and anonymously with an old agent from $B$, and every young agent from $B$ is matched with an old agent from $A$. Equal cohort masses ensure that every agent has exactly two interactions.

Every entrant begins with a clear record. Before actions are chosen, a young agent observes whether the old opponent is clear or stigmatized. Actions are simultaneous. If a player chooses $D$ while the opponent chooses $C$, the defector receives a stigma. For a young defector, that label is observed by the next-period partner. A label acquired by an old agent has no strategic consequence because the old agent exits after the match. No identity or richer history is revealed.

\begin{figure}[t]
\centering
\begin{tikzpicture}[
  >=Latex,
  every node/.style={font=\small},
  role/.style={draw,rounded corners,minimum width=38mm,minimum height=7mm,align=center},
  labelnode/.style={font=\scriptsize,fill=white,inner sep=1pt}
]
\node[role] (YA) at (0,1.2) {Young $A$ at $t$};
\node[role] (OB) at (0,-1.2) {Old $B$ at $t$};
\draw[<->,semithick] (YA)--(OB)
  node[midway,xshift=8mm,labelnode] {match};
\node[role] (OA) at (6.7,1.2) {Old $A$ at $t+1$};
\node[role] (YB) at (6.7,-1.2) {Young $B$ at $t+1$};
\draw[<->,semithick] (OA)--(YB)
  node[midway,xshift=8mm,labelnode] {match};
\draw[->,semithick] (YA.east)--(OA.west);
\node[draw,rounded corners,font=\scriptsize,fill=white,inner sep=2pt]
  (S) at (3.35,2.65) {stigma if $D$ meets $C$};
\draw[->,semithick] (YA.north)
  .. controls +(0,0.7) and +(-0.8,-0.1) .. (S.west);
\end{tikzpicture}
\caption{The directed record. Conduct when young can affect only the action of
the agent's next partner.}
\label{fig:timeline}
\end{figure}

\subsection{Types, information, and payoffs}

Each agent is honest with probability $\pi\in[0,1]$ and strategic with probability $1-\pi$. Types are independent across agents and the common prior $\pi$ is the same in both populations. An honest agent follows a commitment rule: choose $C$ against a clear opponent and $D$ against a stigmatized opponent.

A strategic agent has a private exploitation loss $\ell_i$. It is drawn at entry from the uniform distribution on $[0,1]$ and is observed by agent $i$ before the young-period action is chosen. Thus the decision is made after $\ell_i$ is known. The material payoff to a strategic player is
\[
\begin{array}{c|cc}
 & \text{opponent }D & \text{opponent }C\\
\hline
D & 0 & b\\
C & -\ell_i & 1
\end{array}
\qquad b>1.
\]
In the last interaction, $D$ weakly dominates $C$ and is strictly better for every strategic type with $\ell_i>0$. We select $D$ for the measure-zero zero-loss type when that type is indifferent. Hence every strategic old agent defects. A strategic young agent also defects against a stigmatized old opponent: in equilibrium a stigmatized old agent is known to be strategic and therefore chooses $D$. The same tie-breaking convention applies when $\ell_i=0$. The only nontrivial decision is made by a strategic young agent facing a clear old opponent.

Payoffs from the old interaction are initially undiscounted. Section \ref{sec:discounting} introduces a discount factor. We focus on stationary symmetric equilibria: the two groups use the same strategy and induce the same record distribution at every date.

\begin{table}[t]
\centering
\caption{Terminology}
\label{tab:terms}
\begin{tabular}{p{0.20\textwidth}p{0.70\textwidth}}
\toprule
Term & Definition and role\\
\midrule
Young agent & An entrant in the first of two interactions; the only agent whose current action can affect a future payoff.\\
Old agent & An agent in the final interaction; a strategic old agent always defects.\\
Honest agent & A commitment type who cooperates with a clear opponent and defects against a stigmatized opponent.\\
Strategic agent & A payoff-maximizing type with private exploitation loss $\ell_i$.\\
Clear record & No recorded defection against a cooperator. A clear old agent may be honest or strategic.\\
Stigma & A one-bit record created by choosing $D$ against $C$ and observed by the defector's next partner.\\
Cutoff $x$ & A strategic young agent facing a clear old opponent cooperates if and only if $\ell_i\leq x$.\\
\bottomrule
\end{tabular}
\end{table}

\subsection{Institutional interpretations}

The two dates should be read as two roles or career stages, not necessarily as biological generations. Three interpretations fit the model particularly well.

First, in apprenticeship and professional networks, a junior may work once with a senior from another organization and later become the senior member in a new cross-organizational project. A verified incident of free riding, misappropriation, or nonpayment can be communicated to the next collaborator. The private loss $\ell_i$ captures how exposed the junior would be if trust were not reciprocated.

Second, in family-firm or local supplier networks, a new manager first deals with an incumbent counterpart and later represents the firm in a new match. The record belongs to the manager or firm and affects only the next transaction. The model abstracts from richer repeated relationships in order to isolate that single reputational link.

Third, a recommendation or matching system can implement the directed record. A platform or professional association verifies whether one party exploited a cooperating partner and displays a binary warning to the next recommended counterparty. A clear participant may be recommended for a trust-intensive interaction, whereas a flagged participant receives a defensive response. This is not a literal model of a platform with indefinitely lived sellers and long review histories; it is a benchmark for systems that retain one verified incident for one subsequent match.

\section{Stationary equilibrium}\label{sec:equilibrium}

Suppose strategic young agents use cutoff $x\in[0,1]$ against a clear old opponent. Because losses are uniform, a strategic young agent cooperates with probability $x$ in such a match.

\begin{lemma}\label{lem:beliefs}
Under a stationary cutoff $x$, honest old agents are clear. A strategic old agent is stigmatized with probability
\[
z(x)=\pi(1-x)
\]
and is clear with probability
\[
a(x)=1-z(x)=1-\pi+\pi x.
\]
The total mass of clear old agents and the posterior probability that a clear old agent is honest are, respectively,
\begin{equation}\label{eq:mass-posterior}
m(x)=\pi+(1-\pi)a(x),
\qquad
h(x)=\frac{\pi}{m(x)}.
\end{equation}
A clear old agent is therefore more likely to be honest than an unconditioned agent whenever $\pi\in(0,1)$ and $x<1$.
\end{lemma}

If the focal strategic agent reaches old age with a clear record, the next young opponent cooperates with probability
\begin{equation}\label{eq:q}
q(x)=\pi+(1-\pi)x.
\end{equation}
Honest young agents cooperate with a clear old opponent, and a fraction $x$ of strategic young agents do so. A stigmatized strategic old agent instead faces defection with probability one.

Consider a strategic young agent with loss $\ell$ who observes a clear old opponent. Bayes' rule assigns probability $h(x)$ to the event that the old opponent is honest. Cooperation yields
\begin{equation}\label{eq:UC}
U_C(\ell;x)
=h(x)-(1-h(x))\ell+bq(x),
\end{equation}
whereas defection yields
\begin{equation}\label{eq:UD}
U_D(\ell;x)
=h(x)b+(1-h(x))bq(x).
\end{equation}
If the old opponent is honest, defection gives $b$ now but creates stigma. If the old opponent is strategic, both agents defect, no stigma is created, and the young agent retains the continuation value $bq(x)$.

The payoff difference is
\begin{equation}\label{eq:Delta}
\Delta(\ell;x)
=h(x)\bigl[1-b+bq(x)\bigr]-(1-h(x))\ell.
\end{equation}
It is strictly decreasing in $\ell$. The best-response cutoff to conjecture $x$ is consequently
\begin{equation}\label{eq:T}
T(x)=\left[
\frac{\pi[1-b+bq(x)]}{(1-\pi)a(x)}
\right]_{0}^{1},
\end{equation}
where $[y]_{0}^{1}=\min\{1,\max\{0,y\}\}$. A stationary symmetric Perfect Bayesian equilibrium is a fixed point $x=T(x)$ together with the strategies described above and the Bayes-consistent beliefs in \eqref{eq:mass-posterior}.

For an interior fixed point, define
\begin{equation}\label{eq:g}
\begin{split}
g(x;\pi,b)
&=\pi[1-b+bq(x)]-(1-\pi)a(x)x\\
&=A+Bx-Cx^2,
\end{split}
\end{equation}
where
\begin{equation}\label{eq:ABC}
A=\pi[1-b(1-\pi)],\qquad
B=(1-\pi)[b\pi-(1-\pi)],\qquad
C=\pi(1-\pi).
\end{equation}

\begin{proposition}\label{prop:characterization}
Fix $b>1$ and $\pi\in(0,1)$. A stationary symmetric cutoff equilibrium exists. Moreover:
\begin{enumerate}[label=(\roman*)]
\item $x=0$ is an equilibrium if and only if
\[
\pi\leq 1-\frac{1}{b}.
\]
\item $x=1$ is an equilibrium if and only if $\pi\geq\tfrac12$.
\item Every interior equilibrium is a root of $g(x;\pi,b)=0$. Let
\[
\mathcal D=B^2+4CA.
\]
Whenever $\mathcal D\geq0$, the candidate roots are
\begin{equation}\label{eq:roots}
x_{\pm}
=\frac{B\pm\sqrt{\mathcal D}}{2C};
\end{equation}
an interior root is an equilibrium precisely when it lies in $(0,1)$.
\end{enumerate}
\end{proposition}

The root formula provides a complete characterization for all $b>1$ and $\pi\in(0,1)$. Two regions are especially transparent.

\begin{proposition}
\label{prop:regions}
Let $\varphi=(1+\sqrt5)/2$.\footnote{The bound is exactly the positive root of $b^2-b-1=0$: the sufficiency argument in the proof needs $b^2-b-1\leq0$, and $\varphi$ is where that quadratic first turns positive. The appearance of the golden ratio is this algebraic coincidence and carries no independent economic content.}
\begin{enumerate}[label=(\roman*)]
\item If $1<b\leq\varphi$, the equilibrium is unique. It is $x=0$ when $\pi\leq1-1/b$, the root $x_+$ when $1-1/b<\pi<1/2$, and $x=1$ when $\pi\geq1/2$.
\item If $b>2$ and
\[
\frac12<\pi<1-\frac1b,
\]
there are exactly three equilibria: $x=0$, $x=1$, and the interior root $x_I=x_-$.
\end{enumerate}
\end{proposition}

The bound $\varphi$ is a sufficient global uniqueness condition, not a claim that multiplicity begins sharply at $\varphi$ for every honest share. The root formula shows that additional multiplicity configurations can arise for some $b\in(\varphi,2]$. Proposition~\ref{prop:regions}(ii) isolates the high-temptation region used for the main comparative static.

\section{Cooperation, comparative statics, and stability}
\label{sec:comparative}

We distinguish conditional cooperation from its unconditional frequency. Given cutoff $x$, the probability that a young agent cooperates conditional on meeting a clear old opponent is $q(x)$ in \eqref{eq:q}. The probability that a random old opponent is clear is
\[
m(x)=1-\pi+\pi q(x).
\]
The unconditional rate at which young agents choose $C$ is therefore
\begin{equation}\label{eq:R}
R(x)=m(x)q(x).
\end{equation}
This measure is the endogenous cooperation rate studied below. It does not count the mechanically specified actions of honest old agents.

\begin{proposition}\label{prop:backfire}
Suppose $b>2$ and $\pi\in(1/2,1-1/b)$. Along the interior equilibrium $x_I(\pi)$:
\[
\frac{dx_I}{d\pi}<0,\qquad
\frac{dq(x_I)}{d\pi}<0,\qquad
\frac{dR(x_I)}{d\pi}<0.
\]
Under iterated best responses $x_{n+1}=T(x_n)$, the interior equilibrium is locally unstable. The two strict corner equilibria are locally stable, and $x_I$ is the tipping cutoff separating their basins.
\end{proposition}

The proposition gives a precise sense in which a larger honest share can be associated with less cooperation in equilibrium: both conditional and unconditional young cooperation fall along the interior equilibrium branch. Stability changes the interpretation. Because $x_I$ itself falls, a smaller initial conjecture about strategic cooperation is sufficient to move best responses toward $x=1$. More honesty lowers the coordination hurdle even though cooperation measured \emph{at} the hurdle declines.\footnote{The interior cutoff plays the same role here as the threshold separating basins of attraction in models of equilibrium selection under noisy learning \citep{KandoriMailathRob1993}. We do not pursue a full stochastic-stability analysis---there is no mutation process here---but that literature likewise highlights the role of an unstable equilibrium as a coordination boundary.}

\begin{figure}[t]
\centering
\begin{minipage}[t]{0.48\textwidth}
\centering
\begin{tikzpicture}
\begin{axis}[
  width=\linewidth,
  height=5.3cm,
  xmin=0,xmax=1,ymin=0,ymax=1,
  xlabel={honest share $\pi$},
  ylabel={cutoff $x$},
  samples=100,
  axis lines=left,
  tick label style={font=\scriptsize},
  label style={font=\small}
]
\addplot[blue,thick,domain=0:0.3333] {0};
\addplot[red,thick,domain=0.3334:0.4999] {cutoffroot(x,1.5,1)};
\addplot[blue,thick,domain=0.5:1] {1};
\draw[dashed,gray] (axis cs:0.3333,0)--(axis cs:0.3333,1);
\draw[dashed,gray] (axis cs:0.5,0)--(axis cs:0.5,1);
\end{axis}
\end{tikzpicture}

\smallskip
\small (a) Unique cutoff for $b=1.5$.
\end{minipage}\hfill
\begin{minipage}[t]{0.48\textwidth}
\centering
\begin{tikzpicture}
\begin{axis}[
  width=\linewidth,
  height=5.3cm,
  xmin=0,xmax=1,ymin=0,ymax=1,
  xlabel={honest share $\pi$},
  ylabel={young cooperation rate $R$},
  samples=120,
  axis lines=left,
  axis on top,
  tick label style={font=\scriptsize},
  label style={font=\small},
  legend style={font=\scriptsize,at={(0.03,0.97)},anchor=north west,draw=none}
]
\path[fill=gray!15] (axis cs:0.5,0) rectangle (axis cs:0.6667,1);
\addplot[blue,thick,domain=0:0.6667] {x*(1-x+x^2)};
\addlegendentry{$x=0$}
\addplot[blue,densely dotted,thick,domain=0.5:1] {1};
\addlegendentry{$x=1$}
\addplot[red,thick,domain=0.5001:0.6666] {youngrate(x,3,-1)};
\addlegendentry{$x=x_I$}
\draw[dashed,gray] (axis cs:0.5,0)--(axis cs:0.5,1);
\draw[dashed,gray] (axis cs:0.6667,0)--(axis cs:0.6667,1);
\end{axis}
\end{tikzpicture}

\smallskip
\small (b) Equilibrium branches for $b=3$.
\end{minipage}
\caption{Uniform-loss benchmark. Panel (a) shows the unique equilibrium when temptation is moderate. Panel (b) shades the three-equilibrium region. The interior young-cooperation rate decreases with $\pi$, while the strict corner branches do not.}
\label{fig:comparative}
\end{figure}

Figure~\ref{fig:comparative} illustrates both regimes requested in the interpretation. For $b=1.5$, the unique cutoff moves from complete strategic defection through an interior range to complete strategic cooperation. For $b=3$, the shaded region contains the two corner equilibria and the declining interior tipping branch.

\section{A robust worst-case benchmark}\label{sec:robust}

The baseline uses expected payoffs and the Bayes posterior $h(x)$. As a robust alternative, suppose a strategic young agent evaluates each action at the worst possible current old-opponent type, while taking the future cooperation probability $q(x)$ as given. If the agent defects, the payoff is $b$ against an honest old opponent and $bq(x)$ against a strategic old opponent. Because $q(x)\leq1$, the worst-case payoff is $bq(x)$. If the agent cooperates, the two payoffs are $1+bq(x)$ and $-\ell+bq(x)$, so the worst-case payoff is $-\ell+bq(x)$.

\begin{proposition}\label{prop:maximin}
Under the worst-case rule described above, every strategic young agent with $\ell>0$ defects against a clear old opponent. The robust cutoff is $x=0$.
\end{proposition}

The result is deliberately stark. Bayesian cooperation relies on assigning positive probabilities to the benefits and reputational consequences of each opponent type. A player who instead protects against the worst type action by action regards cooperation as yielding exactly $\ell$ less in the worst case. If the future cooperation probability is itself treated as ambiguous, taking its worst value only reinforces the conclusion.

\section{Record clearing and discounting}\label{sec:forgiveness}

\subsection{Probabilistic record clearing}

Suppose a stigma created when young is cleared before the old-period match with probability $\rho\in[0,1]$. Under a conjectured cutoff $x$, a strategic old agent is clear with probability
\[
a_\rho(x)=1-(1-\rho)\pi(1-x).
\]
Conditional on observing a clear old opponent, the honest posterior becomes
\[
h_\rho(x)=\frac{\pi}{\pi+(1-\pi)a_\rho(x)}.
\]
The mass of clear old agents and the resulting unconditional young-cooperation rate are
\begin{equation}\label{eq:R-rho}
m_\rho(x)=\pi+(1-\pi)a_\rho(x),
\qquad
R_\rho(x)=m_\rho(x)q(x).
\end{equation}

Cooperation still guarantees a clear record. Following defection against an honest old opponent, the continuation value $bq(x)$ is restored with probability $\rho$. The cooperation-defection payoff difference is therefore
\begin{equation}\label{eq:Delta-rho}
\Delta_\rho(\ell;x)
=h_\rho(x)\bigl[1-b+(1-\rho)bq(x)\bigr]
-(1-h_\rho(x))\ell.
\end{equation}
The associated best-response cutoff is
\begin{equation}\label{eq:T-rho}
T_\rho(x)=
\left[
\frac{\pi[1-b+(1-\rho)bq(x)]}
{(1-\pi)a_\rho(x)}
\right]_{0}^{1}.
\end{equation}
For every fixed $x$, $T_\rho(x)$ is weakly decreasing in $\rho$: clearing reduces the reputational cost of defection and also enlarges the pool of clear strategic old agents.

For an interior equilibrium, define
\begin{equation}\label{eq:g-rho}
g_\rho(x)
=\pi[1-b+(1-\rho)bq(x)]
-(1-\pi)a_\rho(x)x.
\end{equation}
The corner conditions are
\begin{align}
x=0\text{ is an equilibrium}
&\Longleftrightarrow
1-b+(1-\rho)b\pi\leq0,\label{eq:rho-zero}\\
x=1\text{ is an equilibrium}
&\Longleftrightarrow
2\pi-1-b\pi\rho\geq0.\label{eq:rho-one}
\end{align}

\begin{proposition}\label{prop:forgiveness}
At a locally stable interior equilibrium, the cutoff and both cooperation measures $q$ and $R_\rho$ strictly decrease with $\rho$. At a locally unstable interior tipping equilibrium, the cutoff increases with $\rho$, so the high-cooperation basin becomes harder to reach. If $\rho=1$, the unique strategic cutoff is $x=0$.
\end{proposition}

Thus the pointwise incentive effect of forgiveness is unambiguous, but a blanket comparison between arbitrarily selected equilibria would not be. Distinguishing stable outcomes from tipping equilibria resolves the apparent ambiguity.

\subsection{Discounting}\label{sec:discounting}

Let $\delta\in[0,1]$ discount the old-period payoff. Current payoffs are not discounted, since the temptation to exploit an honest opponent is realized immediately and only the reputational return to staying clear lies in the future. The payoff difference becomes
\[
\Delta_\delta(\ell;x)
=h(x)\bigl[1-b+\delta bq(x)\bigr]-(1-h(x))\ell,
\]
and an interior equilibrium solves
\[
g_\delta(x)
=\pi[1-b+\delta bq(x)]-(1-\pi)a(x)x=0.
\]

\begin{proposition}\label{prop:discounting}
If $\delta=0$, the unique stationary equilibrium is $x=0$. At every locally stable interior equilibrium, the cutoff is strictly increasing in $\delta$.
\end{proposition}

Lower patience therefore weakens cooperation through the same channel already used for record clearing.\footnote{The comparative-static proof parallels the record-clearing argument, but discounting leaves $a(x)$ unchanged and scales only the continuation payoff. Unlike clearing, it does not change which old agents count as clear.} The result cannot be recovered by replacing $b$ everywhere with $\delta b$: doing so would also discount the immediate payoff from exploiting an honest old opponent, which is realized before the agent has any old age left to discount.

\section{Conclusion}\label{sec:conclusion}

We studied a minimal intergenerational reputation mechanism, deliberately stripped down: agents interact once when young and once when old, a single bit of stigma links the two matches, and no identity or deeper history is ever recorded. Honest agents follow a commitment rule; strategic young agents trade off current exploitation against the value of entering old age with a clear record.

Accounting for the information contained in a clear record is essential. Since honest agents remain clear while some strategic agents do not, the honest probability among clear old agents differs from the population share; ignoring this selection produces beliefs that are inconsistent with the intended Bayesian equilibrium. With that selection incorporated, equilibrium reduces to a quadratic cutoff equation: a simple region has a unique equilibrium, and a high-temptation region has two strict corners coexisting with an interior tipping equilibrium.

What is maintained, and qualified, is the comparative static that motivated the exercise. A larger honest share does lower cooperation along the interior branch---but that branch is unstable, and its cutoff separates the basins of the two corner equilibria rather than describing a locally stable outcome. The right reading is a shift in a coordination threshold, not a verdict that honesty is counterproductive. The robust maximin benchmark is more clear-cut: it eliminates strategic cooperation altogether, since an agent who insures against the worst case treats the exploitation loss as a sure cost with no offsetting sure benefit. Record clearing weakens the incentive to preserve a clear record at every fixed conjecture, and under full clearing strategic agents always defect.

Taken together, the results suggest a qualified design lesson rather than a simple one: even a one-bit, directed record can discipline behavior on its own, without repeated bilateral interaction or community-wide monitoring, but its informational content, its persistence, and its role in equilibrium selection have to be analyzed jointly.\footnote{Minimal records are attractive precisely because they are cheap to build and hard to game, but that same minimalism means small design choices---what gets recorded, for how long, and how beliefs are updated from it---do a disproportionate share of the disciplining work.} A system that clears records to appear forgiving, or that fails to update beliefs about who reaches old age clear, may therefore dissipate the incentive the record was meant to create.

\appendix

\section{Proofs}\label{app:proofs}

\subsection{Proof of Lemma~\ref{lem:beliefs}}

\begin{proof}
Every entrant is clear. An honest young agent cooperates with a clear old opponent. Against a stigmatized old opponent, the honest young agent defects; in equilibrium the stigmatized old agent is strategic and also defects. Hence an honest young agent never defects against a cooperator and remains clear when old.

A strategic young agent becomes stigmatized precisely when two events occur: the old opponent is honest, which has unconditional probability $\pi$, and the strategic loss exceeds $x$, which has probability $1-x$. Thus $z(x)=\pi(1-x)$ and $a(x)=1-z(x)$. The mass of clear old agents is the mass $\pi$ of honest agents plus the mass $(1-\pi)a(x)$ of clear strategic agents. Bayes' rule then gives \eqref{eq:mass-posterior}.
\end{proof}

\subsection{Proof of Proposition~\ref{prop:characterization}}

\begin{proof}
Equation \eqref{eq:Delta} is strictly decreasing in $\ell$, so the best reply has a cutoff. Substituting \eqref{eq:mass-posterior} into the zero of \eqref{eq:Delta} yields the unclipped expression in \eqref{eq:T}. The clipped map $T:[0,1]\to[0,1]$ is continuous, so Brouwer's theorem gives a fixed point.

At $x=0$, the fixed-point condition is $T(0)=0$, or
\[
g(0;\pi,b)=A=\pi[1-b(1-\pi)]\leq0.
\]
For $\pi>0$, this is equivalent to $\pi\leq1-1/b$. At $x=1$, the condition
$T(1)=1$ is equivalent to
\[
g(1;\pi,b)=2\pi-1\geq0.
\]
For an interior fixed point clipping is inactive, so $g(x;\pi,b)=0$. Expanding gives \eqref{eq:g}--\eqref{eq:ABC}; the quadratic formula gives \eqref{eq:roots}. Conversely, every retained root or valid corner satisfies the corresponding best-response condition.
\end{proof}

\subsection{Proof of Proposition~\ref{prop:regions}}

\begin{proof}
The function $g(x)=A+Bx-Cx^2$ is strictly concave because $C=\pi(1-\pi)>0$.

Suppose first that $1<b\leq\varphi$. If $\pi<1/2$ and $A>0$, then $g(0)>0>g(1)$, so strict concavity gives a unique root in $(0,1)$ and neither corner is an equilibrium. If $A\leq0$, then $\pi\leq1-1/b$ and
\[
B=(1-\pi)[(b+1)\pi-1]
\leq
\frac{1-\pi}{b}(b^2-b-1)
\leq0.
\]
Thus $g$ decreases from a nonpositive value and $x=0$ is the only equilibrium. If $\pi\geq1/2$, then $g(1)\geq0$. Moreover, $1-1/b<1/2$ for $b<2$, so $A>0$. Concavity implies that $g$ is positive between its endpoint values (apart from a possible equality at $x=1$), and $x=1$ is the unique equilibrium. This proves part (i).

For part (ii), $b>2$ and $1/2<\pi<1-1/b$ imply $g(0)<0<g(1)$. Hence both corners satisfy their strict best-response inequalities. Strict concavity gives one root in $(0,1)$ and a second root above one. The interior root is the smaller root $x_-$ in \eqref{eq:roots}; no other fixed point exists.
\end{proof}

\subsection{Proof of Proposition~\ref{prop:backfire}}

\begin{proof}
At $x_I=x_-$,
\[
g_x(x_I)=B-2Cx_I=\sqrt{\mathcal D}>0.
\]
Direct differentiation gives
\[
g_\pi(x)
=1+2(1-\pi)x+(2\pi-1)x^2
+b(2\pi-1)(1-x)>0
\]
throughout the three-equilibrium region. Hence
\[
\frac{dx_I}{d\pi}
=-\frac{g_\pi(x_I)}{g_x(x_I)}<0.
\]

Since $q=\pi+(1-\pi)x_I$,
\[
\frac{dq}{d\pi}
=\frac{(1-x_I)g_x-(1-\pi)g_\pi}{g_x}.
\]
Using $g(x_I)=0$, the numerator simplifies to
\[
\frac{1-\pi}{\pi}
\left[
\pi^2x_I^2-(1-\pi+2\pi^2)x_I-\pi(1-\pi)
\right].
\]
The expression in square brackets is a convex quadratic in $x_I$. Its values at zero and one are $-\pi(1-\pi)$ and $-1$, respectively, so it is negative on $[0,1]$. Therefore $dq/d\pi<0$.

The clear mass can be written as $m=1-\pi+\pi q$. Because $q<1$ and $dq/d\pi<0$,
\[
\frac{dm}{d\pi}=-1+q+\pi\frac{dq}{d\pi}<0.
\]
It follows from $R=mq$ that $dR/d\pi<0$.

Finally, at an interior fixed point,
\[
g_x=(1-\pi)a(x)\,[T'(x)-1].
\]
Thus $g_x(x_I)>0$ implies $T'(x_I)>1$, so the interior equilibrium is locally unstable. At each strict corner, clipping makes the best-response map constant in a sufficiently small one-sided neighborhood, establishing local stability. The sign of $g$ changes from negative to positive at $x_I$, so iterated best responses move toward zero below the threshold and toward one above it.
\end{proof}

\subsection{Proof of Proposition~\ref{prop:maximin}}

\begin{proof}
The worst-case payoff from $D$ is $\min\{b,bq(x)\}=bq(x)$. The worst-case payoff from $C$ is $\min\{1+bq(x),-\ell+bq(x)\}=-\ell+bq(x)$. Defection is strictly preferred for every $\ell>0$; the zero-loss type is indifferent and has measure zero.
\end{proof}

\subsection{Proof of Proposition~\ref{prop:forgiveness}}

\begin{proof}
Let $\widehat T_\rho(x)$ denote the unclipped expression in \eqref{eq:T-rho}. Direct differentiation gives
\[
\frac{\partial\widehat T_\rho(x)}{\partial\rho}
=-\frac{\pi[\pi+x(b-\pi)]}
{(1-\pi)a_\rho(x)^2}<0.
\]
Clipping therefore makes $T_\rho(x)$ weakly decreasing in $\rho$.

At an interior equilibrium,
\[
\frac{\partial g_\rho}{\partial\rho}
=-\pi bq(x)-\pi(1-\pi)x(1-x)<0.
\]
Implicit differentiation gives
\[
\frac{dx}{d\rho}
=-\frac{\partial g_\rho/\partial\rho}
{\partial g_\rho/\partial x}.
\]
At a locally stable interior fixed point, $\partial g_\rho/\partial x<0$, so $dx/d\rho<0$. At a locally unstable tipping fixed point, $\partial g_\rho/\partial x>0$, so $dx/d\rho>0$. Since $q$ is strictly increasing in $x$, it decreases with $\rho$ at a stable interior equilibrium.

It remains to account for the direct increase in the clear-record mass in \eqref{eq:R-rho}. Put $k=1-\rho$, $s=1-\pi$, and
\[
H=1-k\pi(1+b-2x).
\]
At a locally stable interior fixed point, $\partial g_\rho/\partial x=-sH<0$, so $H>0$; moreover $H<1$. Implicit differentiation and \eqref{eq:R-rho} yield
\[
\frac{dR_\rho}{d\rho}
=\frac{\pi}{H}\left\{
s(1-x)qH
-[m_\rho+k\pi q]\,[bq+s x(1-x)]
\right\}.
\]
Now $m_\rho>s(1-x)$, because $m_\rho=1-\pi s k(1-x)$ and $s(1-x)(1+\pi k)<1$. Consequently,
\[
b[m_\rho+k\pi q]>s(1-x)H.
\]
The expression in braces is therefore strictly negative, proving that $R_\rho$ decreases with $\rho$ along a stable interior branch.

If $\rho=1$, then
\[
g_1(x)=\pi(1-b)-(1-\pi)x<0
\]
for all $x\in[0,1]$. The only fixed point is therefore $x=0$.
\end{proof}

\subsection{Proof of Proposition~\ref{prop:discounting}}

\enlargethispage{5\baselineskip}

\begin{proof}
At $\delta=0$,
\[
g_\delta(x)=\pi(1-b)-(1-\pi)a(x)x.
\]
Since $b>1$, the first term is negative, and $(1-\pi)a(x)x\geq0$ throughout $[0,1]$. Equivalently, the unclipped best-response cutoff is negative for every conjecture $x$, so the clipped best response satisfies $T_\delta(x)=0$ throughout. Its unique fixed point is therefore $x=0$.

For the comparative static, $\partial g_\delta/\partial\delta=\pi bq(x)>0$. As in the proof of Proposition~\ref{prop:backfire}, $\partial g_\delta/\partial x=(1-\pi)a(x)[T_\delta'(x)-1]$ at an interior fixed point. Local stability therefore implies $\partial g_\delta/\partial x<0$. Implicit differentiation then gives
\[
\frac{dx}{d\delta}
=-\frac{\partial g_\delta/\partial\delta}{\partial g_\delta/\partial x}>0
\]
at every locally stable interior equilibrium.
\end{proof}

\end{document}